\documentclass[structabstract]{aa}  
\RequirePackage{xspace}
\RequirePackage{amsmath}
\RequirePackage{amsfonts}
\RequirePackage{amssymb}

\def\ifundefined#1{\expandafter\ifx\csname#1\endcsname\relax}

\def\la{\mathrel{\hbox{\rlap{\hbox{\lower4pt\hbox{$\sim$}}}\hbox{$<$}}}}
\def\ga{\mathrel{\hbox{\rlap{\hbox{\lower4pt\hbox{$\sim$}}}\hbox{$>$}}}}

\newcommand{\be}{\begin{equation}}
\newcommand{\ee}{\end{equation}}

\newcommand{\bea}{\begin{eqnarray}}
\newcommand{\eea}{\end{eqnarray}}

\ifundefined{ensuremath}\def\ensuremath#1{\relax\ifmmode{#1}}
\else${#1}$\fi\else\relax\fi
\ifundefined{nuc}\def\nuc#1#2{\relax\ifmmode{}^{#1}{\protect\mathrm{#2}}
\else${}^{#1}$#2\fi}\else\relax\fi
\newcommand{\kmps}{\ensuremath{\mathrm{km}~\mathrm{s}^{-1}}\xspace}

\def\Teff{\ensuremath{T_{\mathrm{model}}}\xspace}

\def\alog#1{\times 10^{#1}}

\usepackage{graphicx}
\usepackage{txfonts}
\usepackage{natbib}
\bibpunct{(}{)}{;}{a}{}{,} 
\usepackage{color}
\graphicspath{{./figs/}}

\newcommand{\ie}{i.e.\xspace}
\newcommand{\nhat}{\hat{\mathbf{n}}}
\newcommand{\rin}{\ensuremath{{R_{\mathrm{in}}}} }
\newcommand{\rout}{\ensuremath{{R_{\mathrm{out}}}} }
\newcommand{\Jb}{\ensuremath{{\bar J}}}
\newcommand{\Lb}{\ensuremath{\bar\Lambda}}
\newcommand{\e}{\ensuremath{\epsilon}}
\newcommand{\Jnew}{\ensuremath{{\bar J_{\rm new}}}}
\newcommand{\Jold}{\ensuremath{{\bar J_{\rm old}}}}
\newcommand{\Jfs}{\ensuremath{{\bar J_{\rm fs}}}}
\newcommand{\Snew}{\ensuremath{{S_{\rm new}}}}
\newcommand{\Sold}{\ensuremath{{S_{\mathrm{old}}}}}
\newcommand{\lstar}{\ensuremath{\Lambda^*}}

\begin{document}
   \title{A 3D radiative transfer framework: X. Arbitrary velocity
     fields in the comoving frame}

   \author{          E.~Baron
          \inst{1,2,3,4}
          \and
          Peter H.~Hauschildt
          \inst{1}
          \and
          Bin Chen
          \inst{2}
          \and
          Sebastian Knop
          \inst{1}
          }

   \institute{Hamburger Sternwarte, Gojenbergsweg 112, 21029 Hamburg, Germany\\
              \email{[sknop,yeti]@hs.uni-hamburg.de}
         \and
             Homer L.~Dodge Department of Physics and Astronomy, University of Oklahoma, 440 W Brooks, Rm 100, Norman, OK 73019 USA\\
             \email{[baron,bchen]@ou.edu}
         \and
        Computational Research Division, Lawrence Berkeley
        National Laboratory, MS 50F-1650, 1 Cyclotron Rd, Berkeley, CA
        94720 USA \email{eabaron@lbl.gov}
         \and
        Physics Department, Univ. of California, 
        Berkeley, CA
        94720 USA 
             }
\titlerunning{Arbitrary Velocity Flows}
\authorrunning{E.~Baron et al.}
   \date{}

  \abstract
       {}
     {General 3D astrophysical atmospheres will have random velocity
     fields. We seek to combine the methods we have developed for solving
   the 1D problem with arbitrary flows to those that we have
   developed for solving the fully 3D relativistic radiative transfer
   problem for
 monotonic flows.}
     {The methods developed in the case of 3D atmospheres with
     monotonic flows, solving the fully relativistic problem along
     curves defined by an affine parameter, are very flexible and can
     be extended to the case of arbitrary velocity
     fields in 3D. Simultaneously, the techniques we developed for treating
     the 1D problem with arbitrary velocity fields are easily adapted
     to the 3D problem.}
     {
   The algorithm we present can be used to solve
   3D radiative transfer problems that include arbitrary wavelength
   couplings. We use  
    a quasi-analytic formal solution of the radiative
   transfer equation that significantly improves the overall
   computation speed. We show that the  approximate lambda
   operator developed in previous work gives good convergence, even
   neglecting wavelength coupling. Ng acceleration also gives
   good results. We present tests that are of similar resolution to
   what has been presented using Monte-Carlo techniques, thus our
   methods will be applicable to problems outside of our test setup.
   Additional domain decomposition parallelization strategies will be
   explored in future 
   work. 
   }
     {}

   \keywords{radiative transfer}

   \maketitle
\section{Introduction}
\label{sec:intro}

In a series of papers \citep{hb06,bh07,hb08,hb09a,bhb09,hb10a,shb10}
we have developed a general framework for solving 3D radiative
transfer problems in Cartesian, cylindrical, and spherical coordinates
for both static and monotonic velocity fields in the comoving
frame. We have also developed an Eulerian code for velocities $\la
1000$~\kmps \citep{shb10}.  The neglect of relativistic effects and
resolution constraints limits the
applicability of the Eulerian approach to $v/c \la 0.01$. Our affine
method for solving the fully relativistic transfer equation is exact
to all orders in $v/c$
\citep{bin07,bbh09}. In a parallel series of papers we have introduced
methods for solving the fully relativistic transfer equation in 1D
spherical coordinates with arbitrary velocity fields \citep{bh04,khb09b,khb09}.

\section{Basic formalism}
\label{sec:formalism}

Using the general formalism developed in \citet{bin07} we can derive
the transfer equation in flat spacetime with arbitrary flows. We
choose to work in spherical coordinates without loss of generality.

The photon's four-momentum can be written \be
p^a\equiv\frac{dx^a}{d\xi}=\frac{\rm h}{\lambda_\infty}(1,\hat{\bf
  n}), \ee where $\rm h$ is Planck's constant, $\xi$ is the affine
parameter, $\lambda_\infty$ is the rest frame wavelength, and
$\hat{\bf n}$ is the 3D direction of the photon as seen by a distant,
stationary, observer. The four-velocity of the comoving observer in
an arbitrary flow can be written \be u^a=\gamma({\bf
  r},t)[1,\,\vec{\beta}({\bf r},t)], \ee and the comoving wavelength
$\lambda$ can be obtained using \be \frac{\rm h}{\lambda}=-(u\cdot
p). 
\ee 
Here, $\vec{\beta} = \vec{v}/c$, and $\gamma =
(1-\beta^2)^{-1/2}$, are the usual quantities of special relativity.
The 3D geodesic in the flat spacetime can be parametrized as
\be\label{3D-geo} {\bf{r}}(s)={\bf{r}}_0+\hat{{\bf{n}}}\, s, \ee where
${\bf r}_0$ is the starting point of the characteristics, and $s$ is
the rest frame physical distance related to the affine parameter $\xi$
by \be\label{sdef} s\equiv \frac{\rm h}{\lambda_\infty}\xi .\ee The
radiative transfer equation can be written in terms of the affine
parameter $\xi$ as \citep[see Eq.\,(10) of][]{bin07}: \be\label{step1}
\left.\frac{\partial I_\lambda}{\partial\xi}\right|_\lambda
+\frac{d\lambda}{d\xi}\frac{\partial I_\lambda}{\partial
  \lambda}=-\left(\chi_\lambda
\frac{h}{\lambda}+\frac{5}{\lambda}\frac{d\lambda}{d\xi}\right)I_\lambda+\eta_\lambda
\frac{h}{\lambda}, \ee where $I_\lambda({\bf r}, t;\nhat)$ is the
specific intensity measured by a comoving observer (note that
$I_\lambda \lambda^5$ is observer-independent) at the (global) rest
frame space-time point $(r,\,t)$, toward the rest frame direction
$\nhat,$ and at the comoving wavelength $\lambda$. When expressing
the 7-D phase-space dependence of the comoving specific intensity,
the only comoving variable we used was the wavelength $\lambda$, in
particular, we did not use angles measured  by a comoving observer to
specify the direction of the photons.  The
advantages for this at first sight odd phase-space configuration have been
explained in detail in \cite{bin07}, and we applied this
technique in \citet{bbh09}. 

We can rewrite Eq.~(\ref{step1}) as
\bea\label{step2}
\left.\frac{d(ct)}{d\xi}\frac{1}{c}\frac{\partial I_\lambda}{\partial
    t}\right|_\lambda + \frac{d\vec{r}}{d\xi}\cdot\nabla I_\lambda
&+&\frac{d\lambda}{d\xi}\frac{\partial I_\lambda}{\partial
  \lambda}=\nonumber\\
&&\qquad -\left(\chi_\lambda \frac{h}{\lambda}+\frac{5}{\lambda}\frac{d\lambda}{d\xi}\right)I_\lambda+\eta_\lambda \frac{h}{\lambda},
\eea or equivalently 
\bea\label{step3}
\left.\frac{d(ct)}{d\xi}\frac{1}{c}\frac{\partial I_\lambda}{\partial
    t}\right|_\lambda + \frac{ds}{d\xi}\frac{d\vec{r}}{d s}\cdot\nabla
I_\lambda &+&\frac{ds}{d\xi}\frac{d\lambda}{ds}\frac{\partial
  I_\lambda}{\partial \lambda}=\nonumber\\
&&\qquad -\left(\chi_\lambda \frac{h}{\lambda}+\frac{5}{\lambda}\frac{ds}{d\xi}\frac{d\lambda}{ds}\right)I_\lambda+\eta_\lambda \frac{h}{\lambda}.
\eea
Then using the definition of $s$ from Eq.~(\ref{sdef}) and the fact that
\be
 \frac{d(ct)}{d\xi} = c p^t = \frac{h}{\lambda_\infty},
\ee
we find
\bea\label{trans}
\left.\frac{1}{c}\frac{\partial I_\lambda}{\partial
    t}\right|_\lambda  + \left.\frac{\partial I_\lambda}{\partial
    s}\right|_\lambda +\frac{d\lambda}{ds}\frac{\partial
  I_\lambda}{\partial \lambda}&=&-\left(\chi_\lambda
  \frac{\lambda_\infty}{\lambda}+\frac{5}{\lambda}\frac{d\lambda}{ds}\right)I_\lambda\nonumber\\
&&\qquad+\eta_\lambda \frac{\lambda_\infty}{\lambda}.
\eea 
Eq.~(\ref{trans}) can be put into our standard form:
\be
\left.\frac{1}{c}\frac{\partial I_\lambda}{\partial
    t}\right|_\lambda  + \frac{\partial I_\lambda}{\partial s} + a(s)\frac{\partial}{\partial\lambda}
(\lambda I_\lambda) + 4 a(s)I_{\lambda} = -\chi_\lambda f(s)
I_\lambda + \eta_\lambda f(s),
\label{eqn:phxform2}
\ee where  
\be\label{fs}
f(s)\equiv\frac{\lambda_\infty}{\lambda}=\gamma({\bf r},t)[1-{\bf\hat{n}}\cdot \vec{\beta}({\bf r},t)]
\ee is simply the Doppler factor, and 
\be
a(s)\equiv\frac{1}{\lambda}\frac{d\lambda}{ds}.
\ee 

Using Eqs.\,(\ref{3D-geo}) and (\ref{fs}), $a(s)$ is found to be
\be
a(s)=\frac{1}{1-{\bf\hat{n}}\cdot\vec{\beta}}\left[\frac{d}{ds}({\bf\hat{n}}\cdot{\vec\beta})-\gamma^2\beta(1-{\bf\hat{n}}\cdot\vec{\beta})\frac{d\beta}{ds}\right], 
\ee where $\beta$ is the magnitude of $\vec{\beta},$ and 
\be
\frac{d}{ds}=\frac{1}{c}\frac{\partial}{\partial t}+{\hat{\bf n}}\cdot
\nabla=\frac{1}{c}\frac{\partial}{\partial t}+\frac{\partial}{\partial s}. 
\ee
When we numerically integrate the radiation transfer equation, $a(s)$ can be approximated as
\be\label{numerical-a(s)}
a(s)\approx\frac{\delta({\bf\hat{n}}\cdot{\vec\beta})-\gamma^2\beta(1-{\bf\hat{n}}\cdot\vec{\beta})\,\delta\beta}{\delta
  s (1-{\bf\hat{n}}\cdot\vec{\beta})}, 
\ee  where $\delta s$ is the differential step size (physical
distance) along the characteristics,
$\delta({\bf\hat{n}}\cdot{\vec\beta})$ and $\delta \beta$ are
 the changes of ${\bf\hat{n}}\cdot{\vec\beta}$ and $\beta$, respectively,
when we move one step forward which includes the changes induced by
both time and spatial advances, for instance
\be
\delta \beta=\beta(s_{i+1}, t_{i+1})-\beta(s_i, t_i).
\ee Since few numerical schemes will be able to provide the fully
implicit derivative, $\delta \beta$, will often be obtained for example
by the backward difference
\be
\delta \beta=(\beta(s_{i}, t_{i})-\beta(s_i, t_{i-1}) + (\beta(s_{i},t_i)-\beta(s_{i-1},t_i).
\ee

In the stationary case, both $\beta$ and $f(s)$ are independent of
time and 
specializing Eqs.\,(\ref{3D-geo}) and (\ref{fs}) to that case, $a(s)$ becomes
\bea
a(s)   
&=&\frac{{\bf
    \hat{n}}\cdot\nabla({\bf\hat{n}}\cdot{\vec\beta})-\gamma^2\beta(1-{\bf\hat{n}}\cdot\vec{\beta})({\bf
    \hat{n}}\cdot\nabla\beta)}{1-{\bf\hat{n}}\cdot\vec{\beta}}
\nonumber \\
&=&-({\bf\hat{n}}\cdot\nabla) \ln
(1-{\bf\hat{n}}\cdot{\vec\beta})-\gamma^2\beta({\bf
\hat{n}}\cdot\nabla\beta), \label{eqn:adef}
\eea
where we have used
the fact that along the characteristics, $d/d s$ no longer contains
the time derivative and is thus the directional
derivative operator, that is,  $d/ds=\hat{\bf n}\cdot\nabla$. 
Recall that in the flat spacetime that we are
considering, our characteristics are straight lines for all velocity flows.

In terms of its spherical components,  $\vec{\beta}$ can be written
\be
\vec{\beta}=\beta_r \hat{\bf e}_r +\beta_{\theta} \hat{\bf e}_{\theta}+\beta_{\phi} \hat{\bf e}_{\phi},
\ee where $\hat{\bf e}_r,\hat{\bf e}_\theta, {\bf \hat{e}}_{\phi}$ are
the spherical orthonormal basis vectors at point ${\bf r}(r,\theta,\phi),$ \ie   
\bea\label{spherical-base}
\hat{\bf e}_r&=&(\sin\theta\cos\phi, \sin\theta\sin\phi, \cos\theta),\cr
\hat{\bf e}_\theta&=&(\cos\theta\cos\phi,\cos\theta\sin\phi,-\sin\theta),\cr
{\bf \hat{e}}_{\phi}&=&(-\sin\phi,\cos\phi,0),
\eea and consequently the $\hat{\bf n}\cdot{\vec\beta}$ in Eq.~(\ref{numerical-a(s)}) can be calculated using
\bea
\hat{\bf n}\cdot{\vec\beta}&=&\beta_r \hat{\bf n}\cdot\hat{\bf e}_r +\beta_{\theta}\hat{\bf n}\cdot \hat{\bf e}_{\theta}+\beta_{\phi} \hat{\bf n}\cdot\hat{\bf e}_{\phi}\cr
&\equiv& \beta_r n_r+\beta_{\theta} n_{\theta}+\beta_{\phi} n_{\phi}
\eea (note that along the characteristics, $\hat{\bf n}$ has constant Cartesian components, $n_x,\, n_y,\, n_z,$ but changing spherical components, $n_r,\, n_{\theta},\, n_\phi$). Writing $\hat{\bf n}=(n_x, n_y, n_z),$ the spherical components $n_r,\,n_\theta,n_\phi$ can be easily computed using Eq.\,(\ref{spherical-base}).

\subsection{Comparison with Mihalas}
\label{sec:mihalas}

At first glance comparing Eq~(\ref{eqn:phxform2}) with Eq.~(2.12) of
\citet{mih80} something seems amiss. Like \citet{mih80}, we work in
the frame where spatial coordinates and clocks are measured by an
observer at rest. However, Mihalas' time derivative contains a Doppler
factor, whereas ours does not. Also, our terms on the right-hand side
contain Doppler factors, $f(s)$, whereas those of Mihalas do not. The
discrepency has been noted in passing by \citet{bin07} and arises
because our $s$ is a true distance measured in the observer's
frame, whereas that of Mihalas, $s_M$, contains an extra Doppler factor: 
\[ ds_M = \frac{\lambda_\infty}{\lambda} ds = f(s)ds \] 
(note that $f(s)$ is not a constant along the characteristics, and therefore $s_M$ 
is not related to the physical distance $s$ by a simple affine transformation). 
Thus, we can transform from $s$ to $s_M$ in Eq.~(\ref{trans}) to find
\be\label{eqn:mih_s}
\left. \frac{\lambda}{\lambda_\infty}\frac{1}{c}\frac{\partial I_\lambda}{\partial
    t}\right|_\lambda  + \left.\frac{\partial I_\lambda}{\partial
    s_M}\right|_\lambda +\frac{d\lambda}{ds_M}\frac{\partial
  I_\lambda}{\partial \lambda}=-\left(\chi_\lambda
+\frac{5}{\lambda}\frac{d\lambda}{ds_M}\right)I_\lambda+\eta_\lambda, 
\ee 
which is very similar to the equation of Mihalas, except that the
coefficient multiplying the time derivative term is the inverse
Doppler factor $f(s)^{-1}$, since we are working with $I_\lambda$
instead of $I_\nu$, as did Mihalas. \citet{jhb09} accidentally forgot
to convert the time derivative from $I_\nu$ to $I_\lambda$ and thus
their Eqns.~(24)--(25) have a coefficient of the time derivative with the
Doppler factor in the numerator, rather than in the denominator. Thus,
the time derivative terms in their Eqns.~(24)--(25) should be multiplied by
$f(s)^{-2}$ to give the correct equations. 

\subsection{Nonmonotonic flows}
\label{sec:nm}
We are now in a position to tie together the work of \citet{bh04},
\cite{khb09b} and \citet{bhb09}. The formal solution in the monotonic
case is an initial value problem in wavelength, but in the arbitrary
flow case \emph{both} spatial coordinates and wavelengths are
\emph{fully} coupled.
This poses a 
significant memory cost, since the matrix obtained by
finite-differencing the equations now contains every wavelength and
not just the spatial points along the characteristic. The
computational cost is surprisingly low because
the linear system can be solved using the
semi-analytic method of \citet{khb09b}. While 
the framework that was given in \citet{bh04} and \citet{bhb09} was formulated for just
one  wavelength discretization, we included here the fully implicit
discretization developed in \citet{hb04}.   
Furthermore, we used the new formal solution that avoids negative generalized
opacities \citep{khb09}. 

The stationary equation of radiative transfer in its characteristic form for the specific
intensity $I$ along a path $s$ reads
\begin{equation}
\label{eq:eqrt}
\frac{\mathrm{d} I_l}{\mathrm{d} s} = f(s)\eta_l - f(s)\chi_l I_l - 4 a_l I_l - a_l \frac{\partial(\lambda I_l)}{\partial \lambda} ,
\end{equation}
where $\eta$ is the emissivity, $\chi$ the opacity, and the subscript $l$
indicates dependence on wavelength. The $\frac{\partial}{\partial
  \lambda} $--term  is the coupling term between  
the wavelengths and depends on the structure of the atmosphere and on the
mechanism of the coupling \citep{mih80}.

The wavelength derivative can be discretized in two ways as described in
\cite{hb04}. The different discretizations can be mixed via a
Crank-Nicholson-like scheme with a mixing parameter $\xi \in [0,1]$. The wavelength
discretized equation of radiative transfer can then be written as
\begin{eqnarray}
\label{eq:eqrt2}
\frac{\mathrm{d} I_l}{\mathrm{d} s} &=& f(s)\eta_l - f(s)\chi_l I_l - a_l \left(4 + \xi \;  p_l^| \right) I_l \nonumber \\ 
                                    &{}& \;  - \xi \; a_l \left( p_l^- I_{{l-1}} + p_l^+ I_{{l+1}}  \right)    \nonumber \\
				    &{}& \;  - [1 - \xi] \;  a_l \left(   p_l^- I_{{l-1}} + p_l^| I_l + p_l^+ I_{{l+1}}      \right),
\end{eqnarray}
where the $p_l^\bullet$ coefficients in an ordered wavelength grid $\lambda_{l-1}
< \lambda_l < \lambda_{l+1}$ are defined as
\begin{eqnarray}
\left.
\begin{array}{r c l}
 \displaystyle p_l^- & = & \displaystyle - \frac{\lambda_{l-1}}{\lambda_l - \lambda_{l-1}} \\
 \displaystyle p_l^| & = & \displaystyle \phantom{-}  \frac{\lambda_{l}}{\lambda_l - \lambda_{l-1}}  \\
 \displaystyle p_l^+ & = & \displaystyle \phantom{-} 0
\end{array}
\right\} & \mathrm{\>\>\>for~} & a_{\lambda_l} \ge 0 \\
\left.
\begin{array}{r c l}
 \displaystyle p_l^- & = & \displaystyle \phantom{-} 0 \\
 \displaystyle p_l^| & = & \displaystyle \phantom{-}  \frac{\lambda_{l}}{\lambda_l - \lambda_{l+1}} \\ 
 \displaystyle p_l^+ & = & \displaystyle - \frac{\lambda_{l+1}}{\lambda_l - \lambda_{l+1}}
\end{array}
\right\} & \mathrm{\>\>\>for~} & a_{\lambda_l} < 0. 
\end{eqnarray}
The dependence on the sign of $a_\lambda$ is introduced to define local upwind
schemes  \citep[see][]{bh04}.

After introducing a generalized opacity \citep[see][]{khb09b,khb09}
\begin{equation}
\label{eq:chihat}
\hat{\chi}_l = f(s) \chi_l + \xi \; a_l p_l^| ,
\end{equation}
defining the source functions
\begin{eqnarray}
S_l & = & \frac{\eta_l}{\chi_l} \\
\hat{S}_l & = & \frac{\chi_l}{\hat{\chi}_l} \Bigg\{ f(s) S_l - \xi \; \frac{a_l}{\chi_l} \left( p_l^- I_{{l-1}} + p_l^+ I_{{l+1}} \right)  \Bigg\} \label{eq:srchat} \\
\tilde{S}_l & = & -\, \frac{a_l}{\hat{\chi}_l}  \Bigg\{ [1 - \xi] \;  \left(  p_l^- I_{{l-1}} +  p_l^+ I_{{l+1}}\right) + \left( 4 + [1 - \xi] \; p_l^| \right)  I_l \Bigg\}, \label{eq:srctilde}
\end{eqnarray}
a formal solution of the radiative transfer problem can be
formulated. We used a full characteristic method  throughout the 
atmosphere. The spatial position on a characteristic is then discretized on a
spatial grid. In the following a pair of subscript indices will mark the
position in the spatial grid and in the wavelength grid.  Commonly the spatial
grid is mapped locally onto an optical depth grid via the relation $\mathrm{d} \tau_l =
\hat{\chi}_l \mathrm{d} s$.  The formal solution of the radiative
transfer equation~(\ref{eq:eqrt2}) between two points $s_{i-1}$ and $s_{i}$ on a spatial
grid along the photon path can be written in terms of the optical depth  as follows:
\begin{eqnarray}
\label{eq:frmsl}
I_{i,l} & = & I_{i-1,l} e^{- \Delta \tau} +  \delta \hat{I}_{i,l} +  \delta \tilde{I}_{i,l} \\
\delta \hat{I}_{i,l}   & = &\int_{\tau_{i-1}}^{\tau_{i}} \! \! \!  \hat{S}_l  e^{\tau - \tau_{i}} \mathrm{d} \tau = \alpha_{i,l} \hat{S}_{i-1,l} + \beta_{i,l} \hat{S}_{i,l} + \gamma_{i,l} \hat{S}_{i+1,l} \label{eq:deltahat}\\
\delta \tilde{I}_{i,l} & = &\int_{\tau_{i-1}}^{\tau_{i}} \! \! \!  \tilde{S}_l  e^{\tau - \tau_{i}} \mathrm{d} \tau = \tilde{\alpha}_{i,l} \tilde{S}_{i-1,l} + \tilde{\beta}_{i,l} \tilde{S}_{i,l}, \label{eq:deltatilde}
\end{eqnarray}
with $\Delta \tau = \tau_{i+1,l} - \tau_{i,l}$ and $\tau_{i,l} = \int^{s_i}
\hat{\chi}_l(s) \mathrm{d} s$. The $\alpha$-$\beta$-$\gamma$ coefficients are
described in \citet{ok87} and \citet{phhs392}.  $\delta \tilde{I}_l$ in
Eq.~(\ref{eq:deltatilde}) is linearly interpolated and in general different from
the coefficients in Eq.~(\ref{eq:deltahat}) and is therefore marked
with a tilde. 

Eq.~(\ref{eq:frmsl}) can be written in matrix notation for any given
characteristic:
\begin{equation}
\label{eq:frml_sln_matrix}
\mathbf{I} = \mathbf{A} \cdot \mathbf{I} + \mathbf{\Delta {I}}.
\end{equation}
Here $\mathbf{I}$ is a vector with all intensities, $\mathbf{A}$ is a square matrix
that describes the influence of the different intensities upon each other, and
$\mathbf{\Delta {I}}$ is a vector with the thermal emission and scattering
contribution of the source function.  For a characteristic with $n_i$ spatial
points and $n_l$ points in the wavelength grid, the intensity vector
$\mathbf{I}$ has $n_i \times n_l$ entries.  In the following a superscript of
$k$ will label the characteristic at hand. The components of the matrix
$\mathbf{A}$ from Eq.~(\ref{eq:frml_sln_matrix}) at the spatial point $i$ and the
wavelength point $l$ are given by

\begin{eqnarray}
A^{\mathrm{-},k}_{i,l} & = & - \left( \xi \alpha^k_{i,l} +  [1 - \xi]
  \, \tilde{\alpha}^k_{i,l} \right)
\frac{a^k_{i-1,l}}{\hat{\chi}^k_{i-1,l}}
p^{-,k}_{i-1,l} \label{eq:matrix_coeff_start} \\ B^{\mathrm{-},k}_{i,l} & = & - \left( \xi \beta^k_{i,l} +  [1 - \xi]
  \, \tilde{\beta}^k_{i,l} \right) \frac{a^k_{i,l
  }}{\hat{\chi}^k_{i,l  }}    p^{-,k}_{i,l} \\ C^{\mathrm{-},k}_{i,l} & = &- \xi \gamma^k_{i,l}
\frac{a^k_{i+1,l}}{\hat{\chi}^k_{i+1,l}} p^{-,k}_{i+1,l}\\ A^{\mathrm{\diagdown},k}_{i,l} & = & \! \! \exp{(-\Delta
  \tau^k_{i-1,l})} - \tilde{\alpha}^k_{i,l}
\frac{a^k_{i-1,l}}{\hat{\chi}^k_{i-1,l}} \left[ 4 + [1\!-\! \xi]
  p^{|,k}_{i-1,l} \right] \\ B^{\mathrm{\diagdown},k}_{i,l} & = &- \tilde{\beta}^k_{i,l}
\frac{a^k_{i,l}}{\hat{\chi}^k_{i,l}} \left[ 4 + [1 - \xi] \,
  p^{|,k}_{i,l} \right]\\ C^{\mathrm{\diagdown},k}_{i,l} & = & 0 \\                                 A^{\mathrm{+},k}_{i,l} & = &- \left( \xi \alpha^k_{i,l} +  [1 - \xi]
  \, \tilde{\alpha}^k_{i,l} \right)
\frac{a^k_{i-1,l}}{\hat{\chi}^k_{i-1,l}}
p^{+,k}_{i-1,l} \label{eq:matrix_super1} \\ B^{\mathrm{+},k}_{i,l} & = &- \left( \xi \beta^k_{i,l} +  [1 - \xi] \,
  \tilde{\beta}^k_{i,l} \right) \frac{a^k_{i,l  }}{\hat{\chi}^k_{i,l
  }}    p^{+,k}_{i,l} \label{eq:matrix_super2} \\ C^{\mathrm{+},k}_{i,l} & = &- \xi \gamma^k_{i,l}
\frac{a^k_{i+1,l}}{\hat{\chi}^k_{i+1,l}} p^{+,k}_{i+1,l}. \label{eq:matrix_coeff_end} 
\end{eqnarray}
Following \citet{khb09b}, the naming scheme of the quantities in
Eqs.~(\ref{eq:matrix_coeff_start})--(\ref{eq:matrix_coeff_end}) indicates
with which specific intensity element they 
are associated. For an index pair $i$ and $l$ a $\bullet^{\mathrm{-}}$
superscript refers to an intensity at wavelength $l-1$, a
$\bullet^{\mathrm{\diagdown}}$ superscript to the same wavelength, and
$\bullet^{\mathrm{+}}$ to the next wavelength point $l+1$. The $A, B, C$ terms
refer to the spatial points $i-1, i, i+1$, respectively.  Equations
(\ref{eq:eqrt2})--(\ref{eq:matrix_coeff_end}) are nearly identical to those of
\citet{khb09b} except for the explicit Doppler factor $f(s)$, which
arises because
the photon direction is measured by a distant
stationary observer here rather than by a comoving observer, as was
the case in \citet{khb09b}. We also clarified a problem with
$C^{\mathrm{\diagdown},k}_{i,l}$ that was confusing in \citet{khb09b}.  The formal
solution matrix is therefore identical in form to that shown in Figure~1 of
\citet{khb09b}.

An element of the source function vector $\mathbf{\Delta {I}}$ is
given by \citep{khb09b} 
\begin{equation}
\Delta {I}^k_{i,l}   =  {\alpha}^k_{i,l} \hat{S}^k_{i-1,l} + {\beta}^k_{i,l} \hat{S}^k_{i,l} + {\gamma}^k_{i,l} \hat{S}^k_{i+1,l}. \\
\end{equation} Note that all Doppler factors are explicitly
handled by including them in the opacity as $\chi_l*f(s)$. 

From Eq.~(\ref{eq:frml_sln_matrix}) the solution for the specific intensity for a
given spatial point and wavelength reads
\begin{eqnarray}
I^k_{i,l} & =  & \left(1 - B^{\mathrm{diag},k}_{i,l} \right)^{-1} \cdot \big( \Delta {I}^{k}_{i,l} +  B^{\mathrm{sub},k}_{i,l} I^k_{i,l-1} + B^{\mathrm{super},k}_{i,l} I^k_{i,l+1} \nonumber \\
 & & \quad  + A^{\mathrm{sub},k}_{i,l} I^k_{i-1,l-1} + A^{\mathrm{diag},k}_{i,l} I^k_{i-1,l}+  A^{\mathrm{super},k}_{i,l} I^k_{i-1,l+1} \nonumber \\
 & & \quad  + C^{\mathrm{sub},k}_{i,l} I^k_{i+1,l-1} + \phantom{C^{\mathrm{diag},k}_{i,l} I^k_{i+1,l}+}  C^{\mathrm{super},k}_{i,l} I^k_{i+1,l+1} \big). \label{eq:explicit_frml_sln0} 
\end{eqnarray}

\subsubsection{The operator splitting method}\label{sec:ALI}

 Now that we have the formal solution, the full
  scattering problem can be solved by  using operator splitting.
The mean intensity $J_\lambda$ is obtained from the source function
$S_\lambda$ by a formal solution of the radiative transfer equation,
which is symbolically written 
using the $\Lambda$-operator $\Lambda_\lambda$ as
\begin{equation}
     J_\lambda = \Lambda_\lambda S_\lambda.              \label{frmsol}
\end{equation}
For the transition of a two-level atom, we have
\be
     \Jb = \Lb S,           \label{etla}
\ee
where $\bar J=\int \phi(\lambda) J_\lambda \,d\lambda$, 
$\Lb = \int \phi(\lambda) \Lambda_\lambda \,d\lambda$ with the normalized
line profile $\phi(\lambda)$. 

The $\Lambda$-iteration method, i.e.\  solving Eq.~(\ref{etla}) by a fixed-point
iteration scheme of the form 
\be
   \Jnew = \Lb \Sold , \quad
   \Snew = (1-\e)\Jnew + \e B  ,\label{alisol}
\ee
fails in the case of high optical depths and small $\e$. This is because
the largest eigenvalue of the amplification matrix (for
Doppler-profiles) is approximately  
$\lambda_{\rm max} \approx (1-\e)(1-T^{-1})$, where $T$ is the  optical 
thickness of the medium \citep{mkh75}. For small $\e$ and high $T$, this is very close
to unity and, therefore, the convergence rate of the
$\Lambda$-iteration is very  
poor. A physical description of this effect can be found in 
\citet{mih80}.

The idea of the ALI or operator splitting method is to reduce the 
eigenvalues of the amplification matrix in the iteration scheme
\citep{cannon73}  by 
introducing an approximate \Lb-operator (ALO) $\lstar$
and to split \Lb\ according to
\be
           \Lb = \lstar +(\Lb-\lstar) \label{alodef}
\ee
and rewrite Eq.~(\ref{etla}) as
\be
     \Jnew = \lstar \Snew + (\Lb-\lstar)\Sold. 
\ee
This relation can be written as \citep{hamann87}
\be
    \left[1-\lstar(1-\e)\right]\Jnew = \Jfs - \lstar(1-\e)\Jold, \label{alo1}
\ee
where $\Jfs=\Lb\Sold$. Equation~\ref{alo1} is solved to obtain the new
values of  
$\Jb$, which are then used to compute the new 
source function for the next iteration cycle.

Mathematically, the ALI method belongs to the same family of iterative
methods as the Jacobi or the Gauss-Seidel methods
\citep{golub89:_matrix}. These 
methods have the general form
\be
    M x^{k+1} = Nx^{k} + b
\ee
for the iterative solution of a linear system $Ax=b$ where the system
matrix $A$ is split according to $A=M-N$. For the ALI
method we have $M=1-\lstar(1-\e)$ and, accordingly,
$N=(\Lb-\lstar)(1-\e)$ for the system matrix $A=1-\Lb(1-\e)$. The
convergence of the iterations depends on the spectral radius,
$\rho(G)$, of the iteration matrix $G=M^{-1}N$.  For convergence the
condition $\rho(G)<1$ must be fulfilled, this puts a restriction on
the choice of $\lstar$. In general, the iterations will converge
faster for a smaller spectral radius.  To achieve a significant
improvement compared to the $\Lambda$-iteration, the operator $\lstar$ is
constructed so that the eigenvalues of the iteration matrix $G$ are
much less than unity, resulting in swift convergence. Using
parts of the exact \Lb\ matrix (e.g., its diagonal or a tri-diagonal
form) will optimally reduce the eigenvalues of the 
$G$. The
calculation and the structure of $\lstar$ should be simple to
make the construction of the linear system in Eq.~(\ref{alo1}) fast. For
example, the choice $\lstar=\Lb$ is best in view of the
convergence rate (it is equivalent to a direct solution by matrix inversion)
but the explicit construction of \Lb\ is more 
time-consuming than the construction of a simpler $\lstar$.

In the following discussion we use the notation of \citet{phhs392} and
\citet{hb06}.  The 
basic framework and the methods used for the formal solution and the solution
of the scattering problem via operator splitting are discussed in detail in
\citet{hb06} and will therefore not be repeated here. We have extended
the framework  
to solve line transfer problems with a background continuum. The basic 
approach is similar to that of \cite{phhcas93}. In the simple case of a 
two-level atom with background continuum that we consider here as a test case, 
we use a wavelength grid that covers the profile of the line including the 
surrounding continuum. We then use the wavelength-dependent mean intensities
$J_\lambda$ and approximate $\Lambda$ operators $\lstar$ to compute
the profile-integrated 
line mean intensities $\Jb$ and $\bar\lstar$ via
\[
\Jb = \int \phi(\lambda) J_\lambda\,d\lambda
\]
and 
\[
\bar\lstar = \int \phi(\lambda) \lstar\,d\lambda.
\]
$\Jb$ and $\bar\lstar$ are then used to compute an updated value for 
$\Jb$ and the line source function 
\[ S = (1-\epsilon)\Jb+\epsilon B, \]
where $\epsilon$ is the line thermalization parameter ($0$ for a purely
absorptive line, $1$ for a purely scattering line). $B$ is the Planck 
function, $B_\lambda$, profile averaged over the line 
\[ B = \int \phi(\lambda) B_\lambda\,d\lambda \]
via the standard iteration method 

\be
    \left[1-\lstar(1-\e)\right]\Jnew = \Jfs - \lstar(1-\e)\Jold, \label{alo2}
\ee
where $\Jfs=\bar{\Lb}\Sold$. This equation is solved directly to
obtain the new values of 
$\Jb$, which are then used to compute the new 
source function for the next iteration cycle.

We construct the line \Lb\ directly from the 
wavelength-dependent $\lstar$ generated by the solution of the continuum
transfer problems.

Given the form of Eq.~\ref{eq:explicit_frml_sln0} for the formal solution, the
construction of the $\Lambda^\ast$-operator can proceed exactly as described in
\citet{bh04}. However,  to conserve memory, we have implemented
the $\Lambda^\ast$-operator, retaining the spatial off-diagonal terms,
but \emph{neglecting}  the off-diagonal terms
in wavelength. We still find good convergence at  considerable savings in
memory (see below).

\section{Test calculations}
\label{sec:tests}

In this section we present the results of test calculations we performed
to test the new algorithm in terms of accuracy
by regression testing. We compare these to results of the homologous case
\citep{bhb09}, and to the spherical nonmonotonic cases \citep{bh04}.

The test calculations were performed on  Opteron CPUs running Linux
(Franklin and Hopper at NERSC),  on Intel CPUs (Carver at NERSC, ICE2
at HLRN, and our 
own local Xserve clusters), and on IBM CPUs (PWR-4 and PWR-5). The
code was compiled with Gfortran/gcc/g++, ifort/icc/icpc (versions 11
and 12), and xlf/xlc/xlC and with NAG f95/gcc/g++. Using the varied
compiler suites and CPUs allowed us
to find numerous errors.

\subsection{Regression with monotonic case}
\label{subsec:reg}

\begin{figure}[ht]
  \centering
  \includegraphics[width=\columnwidth]{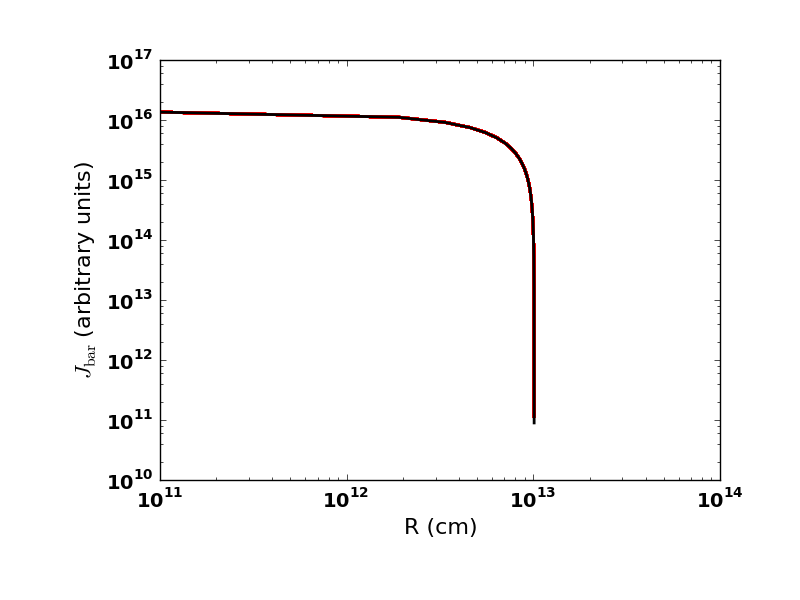}
  \caption{Comparison of monotonic flow case with spherical symmetry
    solved using the full arbitrary velocity field method (red) to the well-tested 1D
    solution (black). The comoving mean intensity is plotted at each
    point in the computational volume. The agreement is at the 1\% level.}
  \label{fig:regression_prof}
\end{figure}

Figure~\ref{fig:regression_prof} shows the profile of the mean intensity
$J$ for homologous flow and spherical symmetry. The test
problem is similar to that of \citet{bhb09}.  These are the basic
model parameters:
\begin{enumerate}
\item An inner radius $\rin=10^{11}\,$cm and an outer radius $\rout =
  1.01\alog{13}\,$cm. 
\item A minimum optical depth in the continuum
  $\tau_{\mathrm{std}}^{\mathrm{min}} =10^{-4}$ and a maximum optical depth in
  the continuum $\tau_{\mathrm{std}}^{\mathrm{max}} = 10^{4}$.
\item A gray temperature structure with $\Teff=10^4$~K. That is the
  temperature solution  of the spherical gray atmosphere problem  with
  effective temperature $\Teff$ \citep{mihalas78sa}. 
\item An outer boundary condition $I_{\rm bc}^{-} \equiv 0$ and an
inner boundary condition $I_\lambda = B_\lambda$ for all
wavelengths.
\item For the initial wavelength the boundary condition is taken from
  that given by the 3D homologous calculation for homologous tests
  and set equal to the Planck function $B_{\lambda_\mathrm{init}}$ for
  non-homologous tests. 
\item A continuum extinction $\chi_c = C/r^2$, with the constant $C$
fixed by the radius and optical depth grids.
\item A parametrized coherent and isotropic continuum scattering given by
\begin{equation}
\chi_c = \epsilon_c \kappa_c + (1-\epsilon_c) \sigma_c
\end{equation}
with $0\le \epsilon_c \le 1$. 
$\kappa_c$ and $\sigma_c$ are the
continuum absorption and scattering coefficients.
 In this work we have neglected scattering
  in the continuum. 
\end{enumerate}

The line of the simple two-level model atom is parameterized by the ratio of the
profile-averaged  line opacity $\chi_l$ to the continuum opacity
$\chi_c$ and 
the line thermalization parameter $\epsilon_l$. For the test cases presented
below, we  used $\epsilon_c=1$ and
the line strength is given by $\Gamma \equiv \chi_l/\chi_c = 10^2$ to simulate a
strong line, with 
varying $\epsilon_l$ (see below). 

\begin{figure}[ht]
  \centering
  \includegraphics[width=\columnwidth]{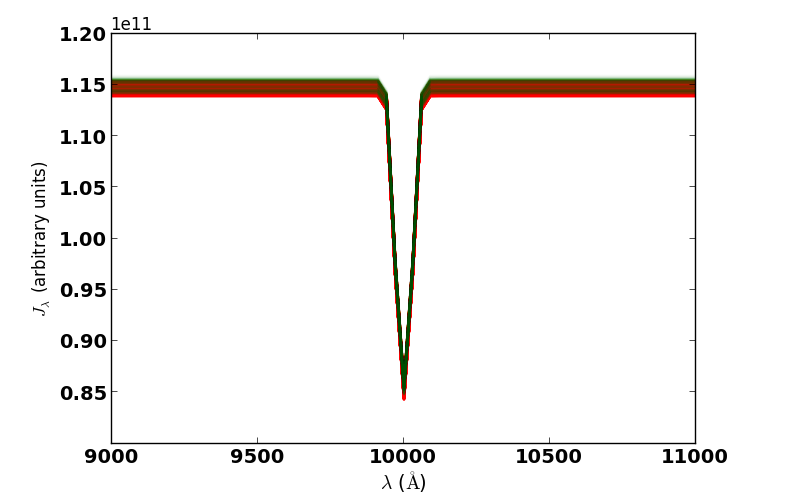}
  \includegraphics[width=\columnwidth]{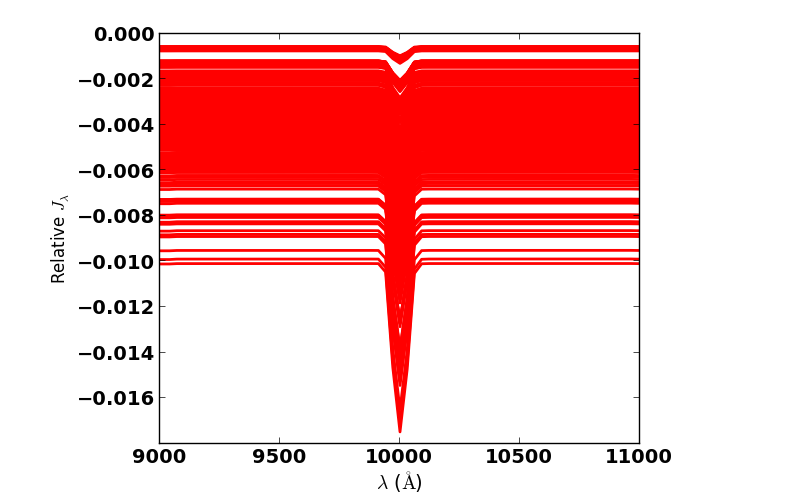}
  \caption{Comparison of monotonic flow case with spherical symmetry
    solved using the full arbitrary velocity field method (green) to
    the 3D monotonic flow solution (red). The comoving mean
    intensity is plotted at the surface of the sphere. The test cases are
    identical, $v_\mathrm{max} = 8 \times 10^4$~\kmps, $129 \times 17
    \times 17$ spatial voxels, 114 wavelength points, and $256\times
    256$ directions. The thermalization parameter in the line is
    $\epsilon = 10^{-3}$ and the line strength is $\Gamma =
      100$. The lower panel shows the relative difference, $\delta J_{\lambda}/J_{\lambda}$, as a
      function of wavelength. 
}
  \label{fig:regression_line}
\end{figure}

The test model is just 
an optically thick sphere put into the 3D grid. The velocity at the
outer radius was set to be relativistic, $v_\mathrm{max} = 8
\times 10^4$~\kmps. The
calculation was performed on a spherical grid with $33^3$ spatial points, $33^2$
solid angle directions, and 22 wavelength points. This and all calculations
presented below were parallelized over characteristics and were run on
parallel clusters. Ng \citep{ng74,auer87} acceleration was used to
significantly speed up the operator splitting iterations for cases
with scattering. 
 Figure~\ref{fig:regression_line}
shows the line profile at the surface, here we compare the 3D
monotonic calculation \citep{bbh07} to the same calculation
using the 3D arbitrary velocity algorithm.  Because of
    the spatial resolution and the way that characteristics end at
    different points in a voxel (as opposed to always ending on a
    radial grid point in the 1D case), it is better to compare 3D
    cases to 3D cases. 

\begin{figure}[ht]
  \centering
  \includegraphics[width=\columnwidth]{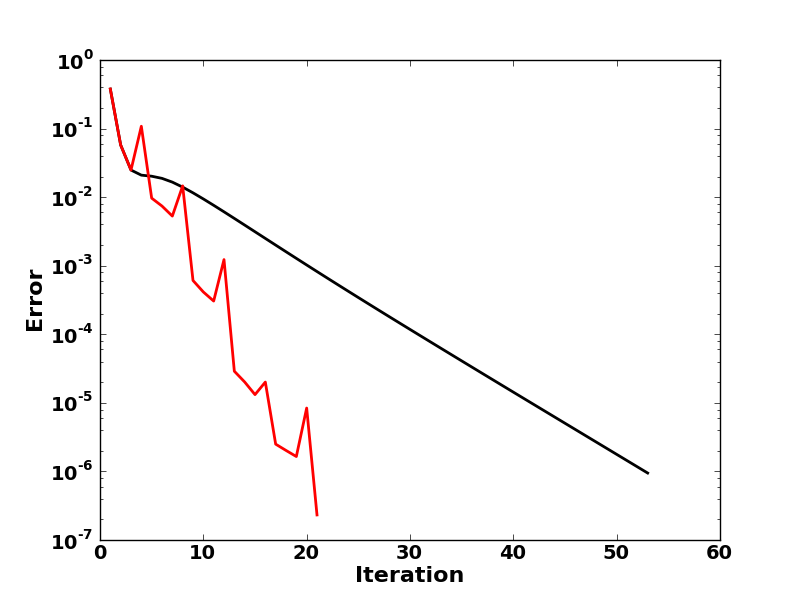}
  \caption{Comparison of the convergence rate of a  monotonic flow
    case with spherical symmetry with the thermalization parameter
    $\epsilon = 10^{-4}$ with and without Ng acceleration. The Ng
    acceleration is identical to that produced by the pure homologous
    3D module.}
  \label{fig:Ng_convergence}
\end{figure}

Figure~\ref{fig:Ng_convergence} shows the convergence rate of a
monotonic flow case, with $\epsilon = 10^{-4}$, with and without Ng
accleration. Not only does Ng acceleration clearly improve the
convergence rate, it gives almost the exact same errors as the
same test setup using both the  3D and 1D homologous algorithms.

\subsection{Sine velocity flow}

\begin{figure}[ht]
  \centering
  \includegraphics[width=\columnwidth]{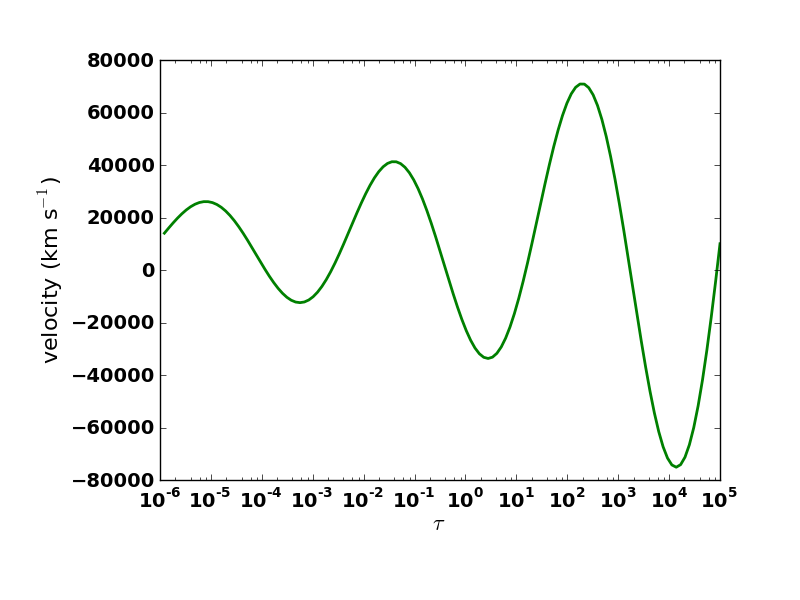}
  \caption{Velocity profile used for the nonmonotonic velocity
    test with a damped sine wave.}
   \label{fig:sin_vel_prof}
\end{figure}

\begin{figure}[ht]
  \centering
  \includegraphics[width=\columnwidth]{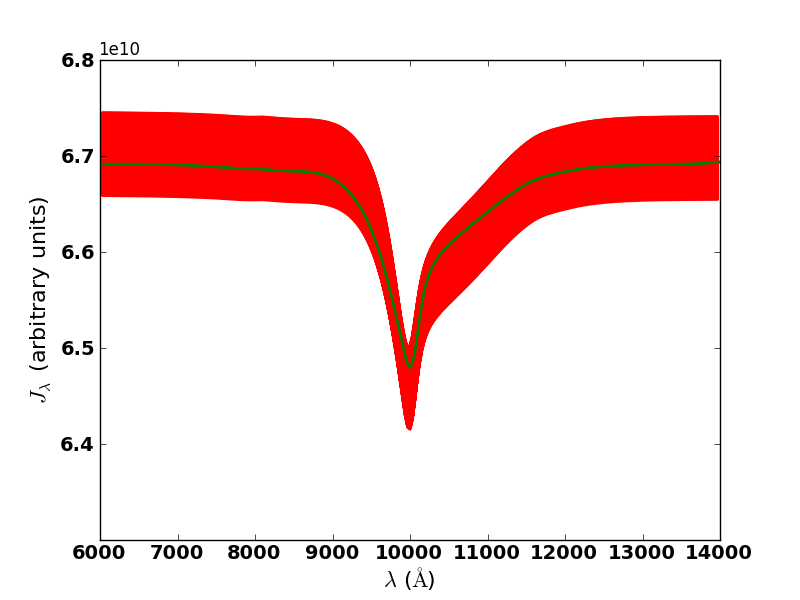}
  \caption{Line profile for the damped sine-wave velocity case with
    high 
    spatial  $n_r = 129$, $n_\theta = 65$, and $n_\phi = 17$, and
    angular resolution $n_\Omega = 512\times512$. 
    The comoving  mean intensity $J_\lambda$ for
    each voxel   on the surface (there are $65\times 17 = 1105$ of
    them) is plotted as a function of
    $\lambda$ (red lines). The green curve is the 1D result using the
    methods of 
    \citet{bh04} and \citet{khb09,khb09b}.  Here the scattering
    fraction is $\epsilon = 10^{-4}$ and the line strength is $\Gamma =
      100$.}
  \label{fig:line_with_sin_HLRN}
\end{figure}

\begin{figure}[ht]
  \centering
  \includegraphics[width=\columnwidth]{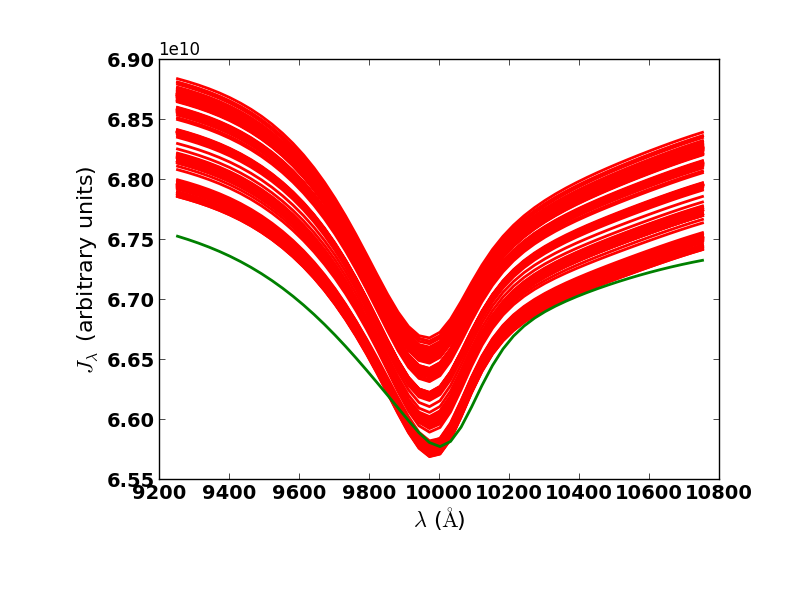}
  \caption{Line profile for the damped sine-wave velocity case with 
    lower
    spatial  $n_r = 129$, $n_\theta = 17$, and $n_\phi = 17$, and
    higher 
    angular resolution $n_\Omega = 512\times512$ than in  
    Figure~\ref{fig:line_with_sin_HLRN}. The comoving mean intensity
    $J_\lambda$ for 
    each voxel   on the surface (there are $65\times 17 = 1105$ of
    them) is plotted as a function of
    $\lambda$ (red lines). The green curve is the 1D result using the
    methods of 
    \citet{bh04} and \citet{khb09,khb09b}.  Here the scattering
    fraction is $\epsilon = 1$  for computational expediency,  and the
    line strength is $\Gamma = 
      100$. With the very low spatial resolution, the 3D result no
      longer envelopes the 1D result (which by assuming spatial
      spherical symmetry has essentially infinite resolution in
      $n_\theta$, and $n_\phi$).} 
  \label{fig:line_with_sin}
\end{figure}

\begin{figure}[ht]
  \centering
  \includegraphics[width=\columnwidth]{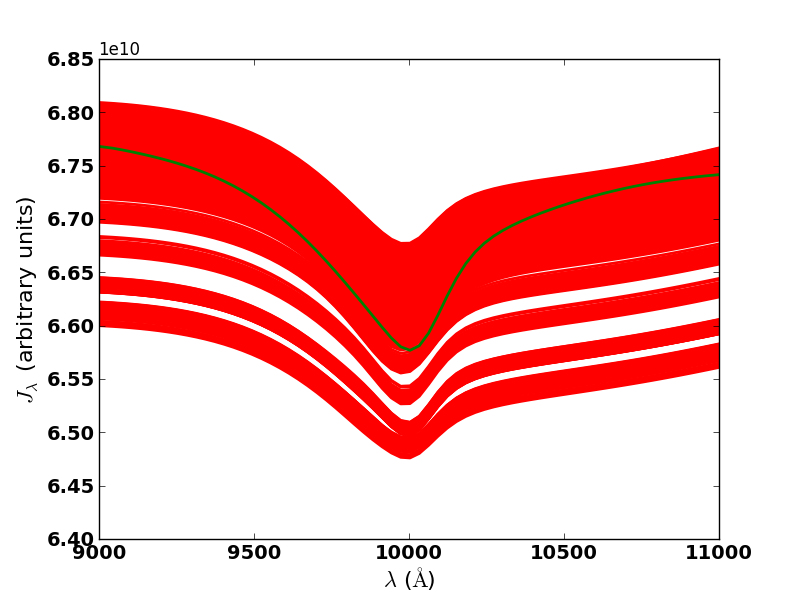}
  \caption{Line profile for the damped sine-wave velocity case with higher
    spatial resolution, but lower angular resolution than in
    Figure~\ref{fig:line_with_sin_HLRN}.
    The  
    spatial resolution is  $n_r = 129$, $n_\theta = 65$, and $n_\phi =
    65$, and   the    
    angular resolution is reduced to  $n_\Omega = 32\times24$. 
    The comoving mean intensity $J_\lambda$ for
    each voxel   on the surface (there are $65\times 17 = 1105$ of
    them) is plotted as a function of
    $\lambda$ (red lines). The green curve is the 1D result using the
    methods of 
    \citet{bh04} and \citet{khb09,khb09b}.  Here the scattering
    fraction is $\epsilon = 1$  for computational expediency,  and the
    line strength is $\Gamma = 
      100$. } 
  \label{fig:line_with_sin_carver}
\end{figure}

Here we again consider a spherically symmetric case with the same
physical parameters as in \S~\ref{subsec:reg}, but the velocity, while
still radial is now given modulated by a damped sine wave.  We have also
introduced a line into the opacity with thermalization parameter
$\epsilon_l = 1$. The thermalization parameter in the continuum
is $\epsilon_c = 1$. This case was
calculated in 1D in \citet{bh04}. Figure~\ref{fig:sin_vel_prof} shows the velocity as a function of
radial optical depth in the continuum.
The spatial grid was $129\times 65\times 17$ and the solid angle
  resolution was set to $512^2$. The maximum velocity is
  $v \approx 10,000$~\kmps and the 
total number of wavelength points is 226.   
  Figure~\ref{fig:line_with_sin_HLRN} shows the comoving mean intensity
  $J_\lambda$ from each surface voxel as a function of wavelength
  $\lambda$. The 1D result is plotted (the green curve) and the agreement is good to
  the sub-percent level. The
variation in the velocity leads to an asymmetric line profile even
in the comoving frame. But since this test has such a high spatial
resolution, it requires a significant amount of memory per process. We
explored the effects of lower spatial resolution, $129\times 17\times
17$, while keeping the high solid angle resolution $512\times 512$. 
Figure~\ref{fig:line_with_sin} shows the  mean intensity as a
function of wavelength for the outmost part of the sphere.  For
computational expediency we set the line thermalization parameter
$\epsilon_l = 1$. The spread
in the results of 1.5 \% is 
indicative of the accuracy of these calculations, whereas the
deviation of the envelope of the solution is indicative 
of the low spatial resolution of this
calculation. While modern nodes may have 24 CPUs, but only about 1 GB
of RAM per CPU, it is unfortunate that the evelope of the low spatial
resolution calculation is offset from the 1D result. This shows that
not only solid angle resolution is  important, but spatial
resolution is quite important as well.
Figure~\ref{fig:line_with_sin_carver} shows the same 
calculation with the spatial grid enhanced to $129\times
65\times 65$ and the
solid angle grid reduced to $32\times 24$ \citep{hb10a}. As can be seen in
Figure~\ref{fig:line_with_sin_carver}, the reduced solid angle
resolution increases the spread by roughly a factor of two, but now
the 3D solution envelopes the 1D nonmonotonic result. 
With this higher spatial resolution, the spread
can be further reduced by increasing the solid angle resolution, which
shows near-perfect weak scaling, and thus does not increase the
wallclock time required for these calculations (although it does
require more CPUs). 

\subsection{Example of radial nonhomologous flows}

The previous tests were all totally spherically symmetric, with
radial variations in the velocity field. 
We now assume the flow to be radial and azimuthally symmetric, \ie
\be
\vec{\beta}({\bf r})=p(\theta)q(r) \hat{\bf e}_r. \label{eqn:nrbetdef}
\ee For this case, we have 
\be 
{\bf \hat{n}}\cdot \vec{\beta}=pqn_r, 
\label{eqn:nr_nhatdotbeta}
\ee
\be
(\hat{\bf n}\cdot\nabla)\beta=pn_rq'(r)+\frac{1}{r}p'(\theta)q n_\theta, 
\label{eqn:nr_nhatdotgradbeta}
\ee
and 
\be
\hat{\bf n}\cdot\nabla(\hat{\bf
  n}\cdot\vec{\beta})=\frac{1}{r}[p'qn_rn_\theta+p(q'r)n_r^2+pq(1-n_r^2)], 
\label{eqn:nr_nhatdotgradnhatdotbeta}
\ee where we made use of 
\be
\dot{r}=n_r,\>\dot{\theta}=\frac{n_\theta}{r},\>\dot{\phi}=\frac{n_\phi}{r \sin\theta},
\ee and
\be
\dot{n}_r=n_\theta\dot{\theta}+\sin\theta n_\phi\dot{\phi}.
\ee 
Here an over dot  implies $\frac{d}{ds},$ e.g.,
$\dot{r}\equiv\frac{dr}{ds},$ etc., 
and a prime denotes differentiation with respect to $\mu = \cos\theta$.

Inserting Eqs.~(\ref{eqn:nr_nhatdotbeta})--(\ref{eqn:nr_nhatdotgradnhatdotbeta})
into Eq.~(\ref{eqn:adef}), 
we find an analytic expression for $a(s)$: 
\bea
a(s)&=&\frac{p'qn_rn_\theta+p(q'r)n_r^2+pq(1-n_r^2)}{r(1-p q
  n_r)}\nonumber\\
&&\qquad -\gamma^2 pq \left[pn_rq'(r)+\frac{1}{r}p'(\theta)q n_\theta \right],
\eea where the spherical components of the unit vector $\hat{\bf n}=(\sin\theta_n \cos\phi_n, \sin\theta_n\sin\phi_n, \cos\theta_n)$ are 
\bea
n_r&=&\sin\theta\sin\theta_n\cos(\phi-\phi_n)+\cos\theta\cos\theta_n,\nonumber\\
n_{\theta}&=&\cos\theta\sin\theta_n\cos(\phi-\phi_n)-\sin\theta\cos\theta_n,\nonumber\\
n_\phi &=&\cos\theta_n.
\eea 

For $p(\theta),$ we use an expansion in terms of
Legendre polynomials (which form a complete and orthonormal basis for 
axially symmetric spherical functions) 
\be
p(\theta)=\sum_{n=0}^{k}{c_n{\cal P}_{n}(\theta)},
\ee with ${\cal P}_0(\mu)=1,$ ${\cal P}_1(\mu)=\mu,$  and ${\cal P}_2(\mu)=\frac{1}{2}(3\mu^2-1),$  etc., where $\mu=\cos\theta.$  We consequently obtain 
\be
p'(\theta)=\sum_{n=1}^{k}{c_n {\cal P}'_{n}(\theta)}
\ee with ${\cal P}_1'=-\sin\theta,$ and ${\cal P}_2'=-\frac{3}{2}\sin
2\theta,$ etc. (note that the case $c_n=c_0\delta^0_n$  degenerates to
homologous flow). We can use this finite expansion in terms of
the Legendre polynomial ${\cal P}_{n}(\theta)$ to construct azimuthally
symmetric jets.  

\begin{figure}[ht]
 \centering
 \includegraphics[width=\columnwidth]{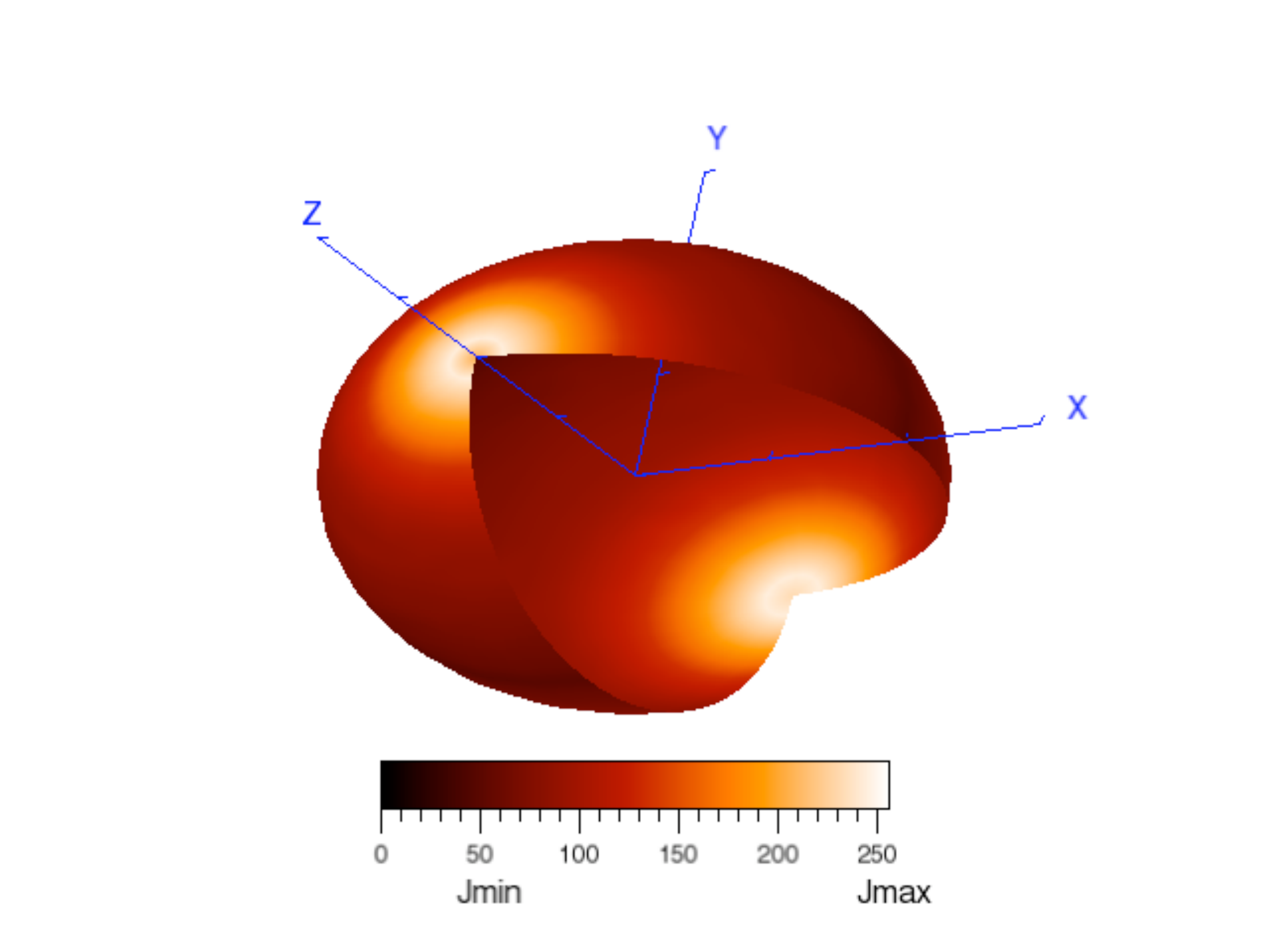}
 \caption{Comoving mean intensity $J(\theta,\phi)$ plotted at the surface
   $R=R_{\rm out}$ for case where $\vec{\bf
     \beta}=c_0rp(\theta)\hat{\bf e}_r,$ and
   $p(\theta)=1+\frac{1}{2}{\cal
     P}_2(\mu)=1+\frac{1}{2}(\frac{3}{2}\cos^2\theta-\frac{1}{2}).$
   The velocity flow is radial, but not spherically symmetric
 to approximate a jet-like flow in the $\pm z$ direction.}
 \label{fig:L2_0p5}
\end{figure}

\begin{figure}[ht]
 \centering
 \includegraphics[width=\columnwidth]{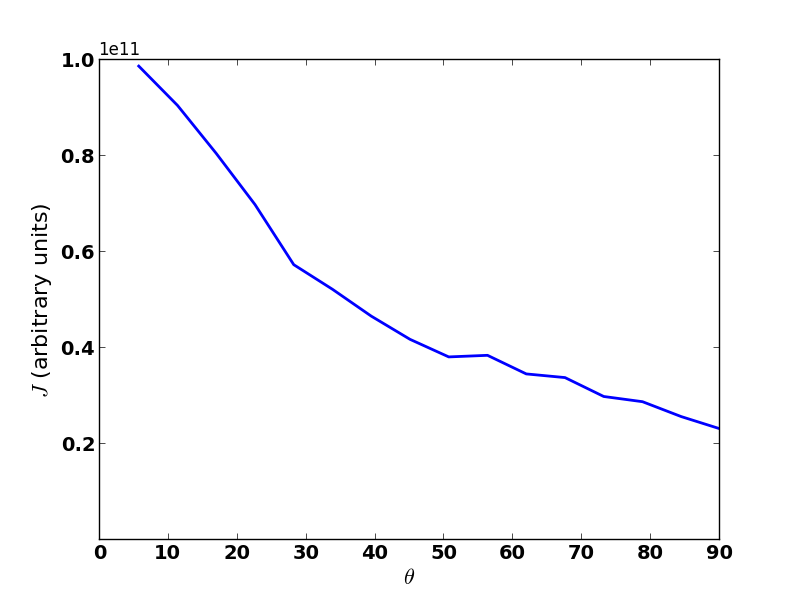}
 \caption{Comoving mean intensity $J(\theta,\phi_0)$ plotted at the surface
   $R=R_{\rm out}$ for a fixed (arbitrarily chosen) value of $\phi =
   \phi_0$. The variation of nearly an order of magnitude from the
   pole to the equator is due to the fact the the continuum level is a
   function of the velocity. }
 \label{fig:L2_0p5_long_slice}
\end{figure}

As a simple example of nonspherically symmetric flow, we run a test case where 
\be\label{beta_axial}
\vec{\bf \beta}=c_0rp(\theta)\hat{\bf e}_r,
\ee  \ie  we take  $q(r)= c_0r$ ($c_0$ is a constant), which simplifies
$a(s)$ to be  
\be
a(s)=c_0\left[\frac{p+p'n_rn_\theta}{1-rn_rp}-\gamma^2\beta(pn_r+p'n_\theta)\right].
\ee Furthermore, we assume
\be\label{L2}
p(\theta)=1+\frac{1}{2}{\cal P}_2(\mu)=1+\frac{1}{2}\left(\frac{3}{2}\cos^2\theta-\frac{1}{2}\right),
\ee \ie, we perturb the homologous flow by adding a second-order
Legendre polynomial with perturbation coefficient $0.5.$ We show the
resulting  mean intensity plot $J(\theta,\phi)$ at the boundary
$R=R_{\rm out}$ in Fig.~\ref{fig:L2_0p5}. We obtained perfect
azimuthal symmetry as expected from the form of $\vec\beta$ in
Eq. (\ref{beta_axial}), although it was not explicitly enforced.  Furthermore, we also recovered the symmetry
with respect to the north/south pole (\ie, symmetry under a reflection
with respect to the equatorial plane,   $\theta\rightarrow
\pi-\theta$), which is characteristic for Legendre polynomials of even
order, see Eq.\, (\ref{L2}). Figure~\ref{fig:L2_0p5_long_slice} shows
a longitudinal slice of the comoving intensity, which shows that the
comoving mean intensity varies by roughly a factor of ten from the
pole to the equator. This is just due to the effect that  the
continuum level varies with the maximum velocity \citep{bhb09}.

\subsection{Checkpointing}

We implemented a checkpointing scheme that allows for 
restarts and new starts with higher resolution in solid angle. We
simply write out (using stream I/O) the $\lstar$ operator once it
has been calculated, and the value of $J$ at each voxel after each ALO
iteration. Since these calculations require significant numbers of
processors, which may go down, or calculations may run out of time, this
allows us to perform perfect restarts. We checked that restarting
from the checkpoint files works perfectly and that the calculations are
converged in a single iteration when restarting from a converged
checkpoint file. We write $\lstar$ and $J$ to separate files, since
that allows us to read just $J$, and then recalculate $\lstar$ if we
restart with a higher resolution. We checked this, and it allows us
to use a low-resolution calculation (performed on fewer
processors, for example) and then converge the higher resolution
calculation 
with roughly half as many iterations for each resolution
increase. Table~\ref{tab:convergence} shows the rate of convergence
for 
a test with homologous velocity fields (chosen for computational
expedience), $v_{\mathrm{max}} = 10^5$~\kmps and $\epsilon =
10^{-5}$. While the restart result still requires several iterations,
this scheme allows for some speedup. The small $\epsilon$ and high
maximum velocity make this test particularly demanding. The
calculations were all 
    restarted such that Ng acceleration cannot begin until the fourth
    iteration, and thus the restart calculations 
    converge more slowly than they could. The scheme could indeed
    possibly be made more efficient by keeping the previous three values of
    $\Jb$, so that the restart iteration could immediately use Ng accleration. 

\begin{table}[ht]
  \centering

  \caption{Convergence rate for homologous test with $v_{\mathrm{max}} = 10^5$~\kmps and $\epsilon =
    10^{-5}$, starting from the previous test. The first test with
    $8\times8$ solid angles is converged from the beginning. The spatial
    resolution is held fixed and $\lstar$ is recalculated at the
    beginning of each new resolution run. }
  \begin{tabular}{ll}
    $\Omega$& Number of iterations\\\hline
    $8\times8$& 29\\
    $64\times64$& 21\\
    $128\times128$& 15\\
    $256\times256$& 13\\
  \end{tabular}
\label{tab:convergence}
\end{table}

\section{Conclusion}
\label{sec:conclusion}

We presented algorithm strategies and details for the solution of the radiative
transfer problem in 3D atmospheres with arbitrary wavelength
couplings.

Future possible applications are the velocity profiles of cool stellar winds,
treatming partial redistribution, calculating  radiative
transfer in shock fronts like in accretion shocks,  calculating
quasar jets, and general relativistic situations such as a rotating
black hole.

\begin{acknowledgements} This work was supported in part  by SFB 676,
  GRK 1354 from the
DFG,   NSF grant AST-0707704, and US DOE Grant
DE-FG02-07ER41517. 
Support for Program number HST-GO-12298.05-A was provided by NASA through 
a grant from the Space Telescope Science Institute, which is operated by the 
Association of Universities for Research in Astronomy, Incorporated, under 
NASA contract NAS5-26555. BC thanks the University of Okalahoma Foundation
for a fellowship.   This research used resources of the National Energy
Research Scientific Computing Center (NERSC), which is supported by the Office
of Science of the U.S.  Department of Energy under Contract No.
DE-AC02-05CH11231; and the H\"ochstleistungs Rechenzentrum Nord (HLRN).  We
thank both these institutions for a generous allocation of computer time.
\end{acknowledgements}

\clearpage
\bibliography{apj-jour,mystrings,refs,baron,sn1bc,sn1a,sn87a,snii,stars,rte,cosmology,gals,agn,atomdata,crossrefs}

\end{document}